**Ancient Luoyang Bridge reveals a simple metagrating model in optics**

Shiming Chen [1], Yangyang Zhou [1], Zhenyu Wang [2] and Huanyang Chen [1, *]

[1] Institute of Electromagnetics and Acoustics and Department of Electronic Science, Xiamen University, Xiamen 361005, China

[2] College of Civil Engineering and Architecture, Zhejiang University, Hangzhou 310058, China

**Abstract:** The ancient Chinese bridge, Luoyang Bridge, has been revealed to obey similar laws to diminish waves, like an optical model, metagrating. Numerical simulations have been performed to verify this finding.

There is ancient bridge in Quanzhou, southern China, which has been exist for more than one thousand years since the Song Dynasty (a picture in Fig 1) [1]. This bridge has benefited people from suffering the flood, and is famous together with another bridge call "Zhaozhou Bridge" in northern China [2]. It is purely a stone bridge with amazing constructing process at that time. Even from modern bridge constructing theory, it has several advantages. Firstly, it consists of many boat-shape pillars, which can not only ease the constructing process, but also reduce the drag from waters. Moreover, the ancient Chinese raised oysters at the pillars, whose shells can reinforce the structures by attaching and merging the stones into one, which is indeed a novel method even from modern biology engineering technology [3]. This bridge has contributed to Quanzhou city, especially the application from being a world heritage.

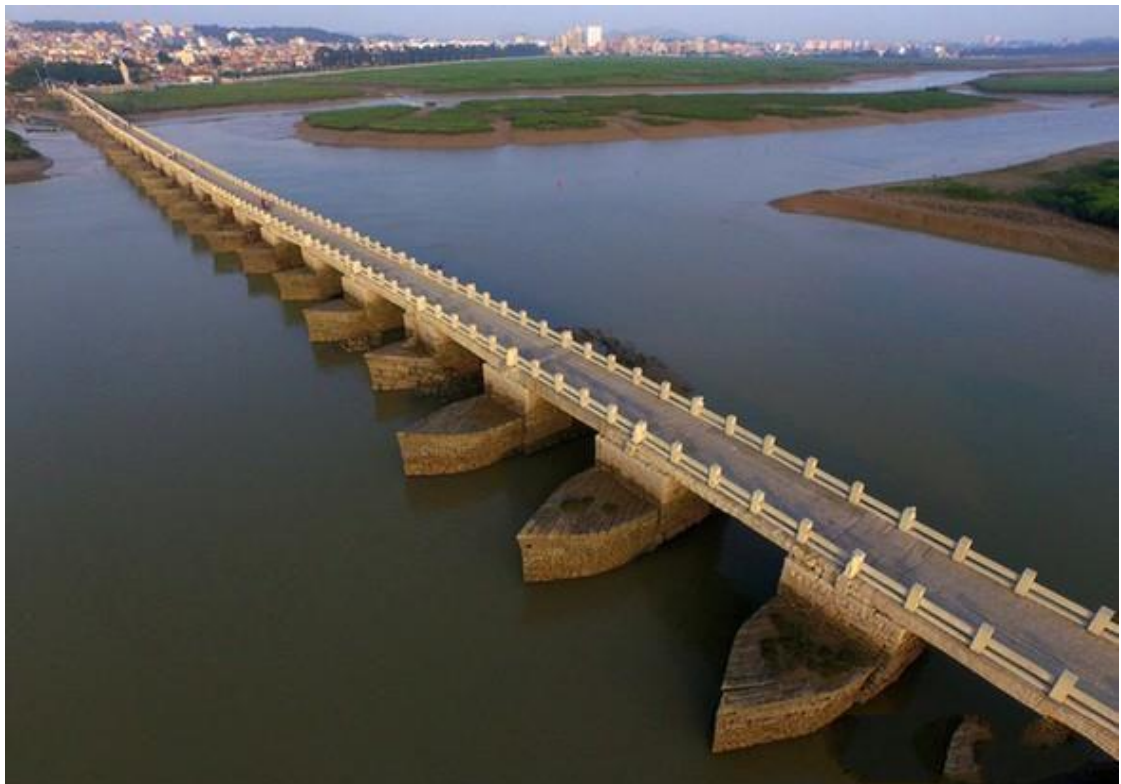

Fig. 1.   Luoyang Bridge [1].

---

*  kenyon@xmu.edu.cn

In this letter, we want to borrow the concept of metamaterials [4] from optics to reveal how these unique boat-shape pillars are so efficient to reduce water waves. Metameterials are artificial materials, whose electrometric parameters could be custom designed, and have been used to implement invisibility cloaks [5] and energy concentrators [6], especially after the proposal of transformation optics [7, 8]. Such a method has been extended to water waves, for designing intriguing devices for ocean waves near the seashore [9, 10]. Therefore, it would be also beneficial for exploring new hydraulic structures design learning from this ancient bridge. We will find that these boat-shape pillars are very similar to a metagrating structure to reduce waves [11].

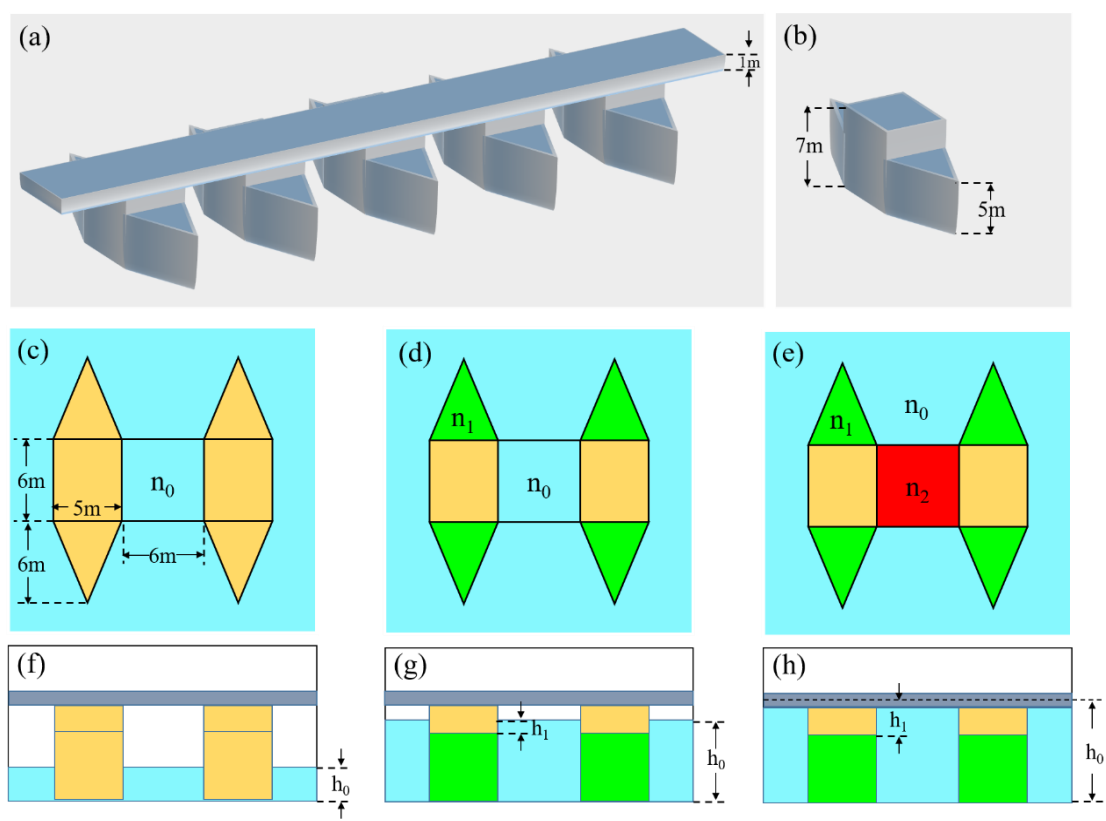

Fig. 2. The schematic diagram of Louyang Bridge, the top view and the side view. (a) Structure of Luoyang Bridge. (b) A single pillar and its size. (c) and (f) are the top view and side view for Case 1 with the water level $h_0$=2.5m. (d) and (g) are the top view and side view for Case 2 with the water level $h_0$=6m and $h_1$=1m. (e) and (h) are the top view and side view for Case 3 with the water level $h_0$=7.5m and $h_1$=2.5m.

The ancient bridge is more than 1000 meters, with 46 pillars [3]. The size of each part is not exactly the same. Therefore, we simply take an approximate size for consideration. The width of the bride is about 6 meters, with a thickness about 1 meter for the top bridge slab. The height of the pillar is about 7 meters, and the boat part is about 5 meter height. The two ends of the boat part are simply taken as triangles with their bottom length 5 meters (which is also the width of the pillar) and their height 6 meters (for the triangular cross section). The distance between two pillars is 6 meters. Please see the schematic plot in Fig. 2 (a), (b) and (c). When the water level $h_0$ is less than 5 meters, the whole pillar works as a perfect reflector (Fig. 2(c) and (f)), and the equivalent refractive index of the background $n_0$ can be chosen as 1, which we denote as Case 1. However, when the water level $h_0$ is larger than 5 meters, the center part of the pillar can still be regarded as a perfect reflector (with a rectangular cross section). But the two triangular ends will have different water levels $h_1$ (Fig. 2 (d) and (g)). If the background refractive index is set as $n_0$=1, the index at the two ends should be $n_1 = \sqrt{h_0/h_1}$[9]. We set it as Case 2. If the water level $h_0$ is larger than 7 meters, but still less than 8 meters (which means that the bridge is almost been covered by the waters), as shown in Fig. 2 (e) and (h), the index at the two ends can still be written as $n_1 = \sqrt{h_0/h_1}$, while the bridge part between each two pillars is no longer with an index of $n_0$, but with a new refractive index $n_2 = 0$. The readers can refer it in [12]. We set it as Case 3. We will numerically explore these three cases for various wavelengths and examine the effect for the bridge to reduce water waves.

Firstly, let us consider a wavelength of 3 meters in Fig. 3. The simulation is only for the metagrating with 10 pillars with a point source at (52m, -15m) and with an amplitude of 1 meter. We then plot the amplitude at y = 20 m for $h_0$ = 2.5, 6, and 7.5 meters, which shows that for Case 2 and 3, when the water level is higher, the bridge can reduce the wave amplitude, when compared to Case 1, as shown in Fig. 3(a). We also plot field patterns for each case in Fig. 3(b), (c) and (d) for Case 1, 2, and 3, respectively. For Case 1, the wave amplitude could already be reduced to certain value and the field is concentrated at the parts between two pillars. However, for Case 2 and 3, the water waves will be concentrated at the triangular ends with higher indexes, and the transmission is greatly reduced. In addition, for Case 3, the zero index part will only allow the normal incident wave component to pass through, which can further reduce the wave transmission.

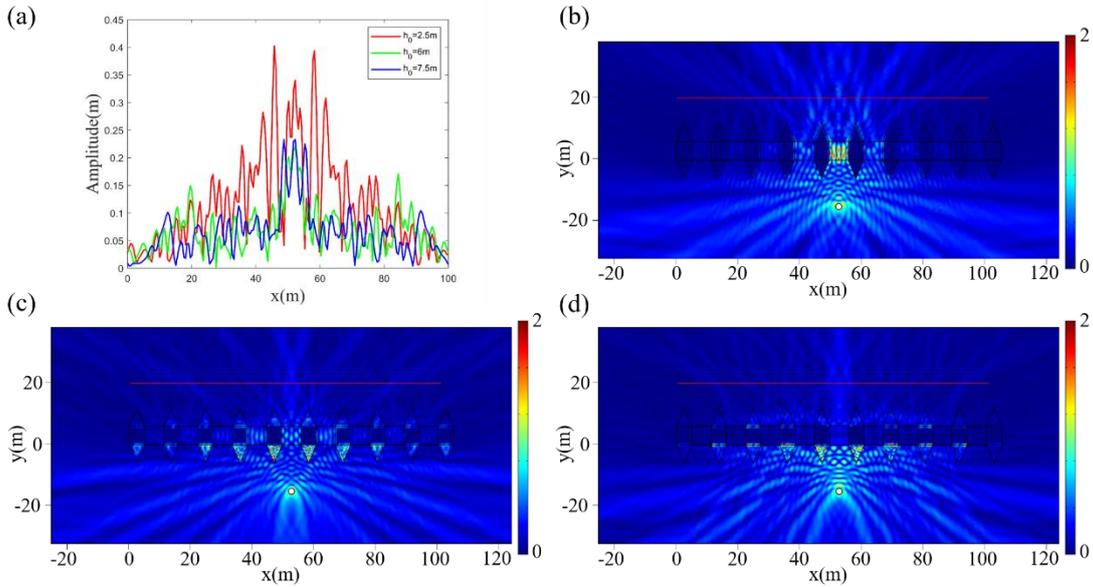

Fig. 3. Wave amplitude with a point source at (52m, -15 m). (a) The amplitude at y = 20 m for different water level $h_0$ = 2.5, 6, and 7.5 meters with the incident wavelength of 3m. (b), (c) and (d) are the field patterns for Case 1, 2, and 3, respectively.

The effect becomes more manifested for longer wavelengths, which are usually more harmful.

For example, we consider a wavelength of 6 meters in Fig. 4. Again, we plot the transmission profiles at the other side when a point source is excited for the three cases in Fig. 4 (a). The perfect reflector grating can reduce the waves to about 20% (Case 1), while the boat shaped pillars in Case 2 can reduce the waves to about 10% and the zero index part in Case 3 can further reduce the waves lower to 5%. We also plot the field patterns for each case in Fig. 4 (b), (c), and (d). The concentrating energy at the parts between pillars for Case 1 is shown in Fig. 4(b), while the concentrating at triangular parts for Case 2 is also clearly shown in Fig. 4(c). Moreover, total reflection for the zero index part is very helpful to further reduce the waves for Case 3, as shown in Fig. 4(d). Similar results for a wavelength of 9 meters are shown in Fig. 5, with (a) the wave amplitude for comparison with different cases; (b), (c), and (d) the corresponding field patterns for Case 1, 2, and 3, respectively.

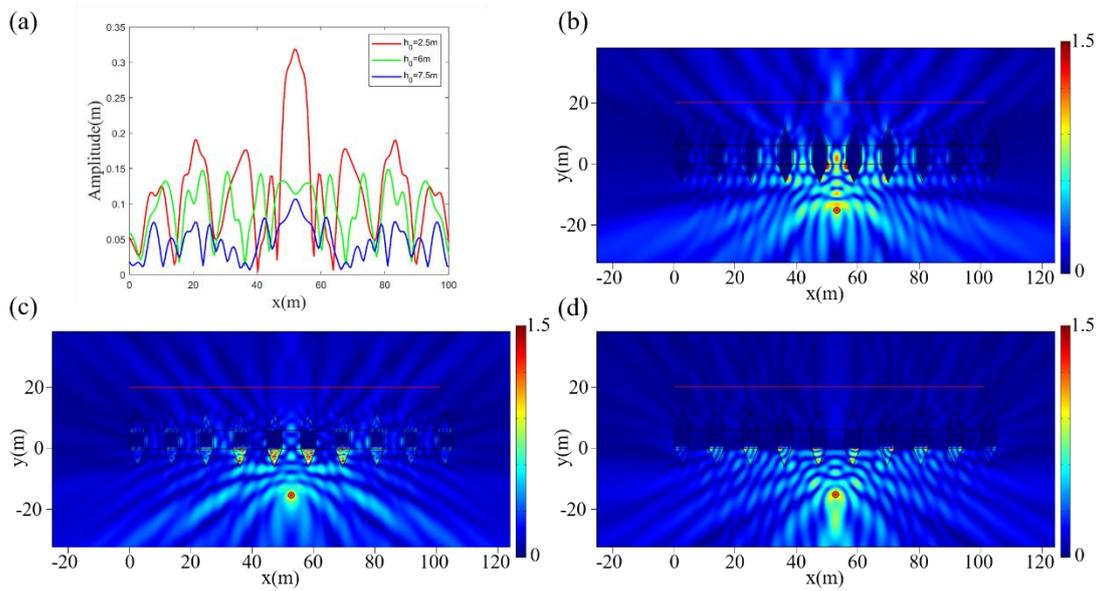

Fig. 4. Wave amplitude with a point source at (52m, -15 m). (a) The amplitude at y = 20 m for different water level $h_0$ = 2.5, 6, and 7.5 meters with the incident wavelength of 6m. (b), (c) and (d) are the field patterns for Case 1, 2, and 3, respectively.

We also examine the transmission profiles for various water levels for different wavelengths (3 meters, 6 meters, and 9 meters) in Fig. 6. From 6(a) to (f), we plot the profiles for $h_0$ = 2.5, 5.5, 6, 6.5, 7, and 7.5m. It clearly shown that, when the water level increases, the transmission will be reduced from about 20% to about 5%, as also revealed in the previous section. Hence, the results are very robust. In fact, we did not consider the loss in simulations. As the wave will be focused at the triangular ends with stronger amplitude, which will be easy to dissipate as the triangular platforms have smaller depth and the surface wave will be diminished when hitting the surface. Hence, the triangular ends work as a perfect match layer [13] such that more waves will be dissipated, and the transmission would be even smaller.

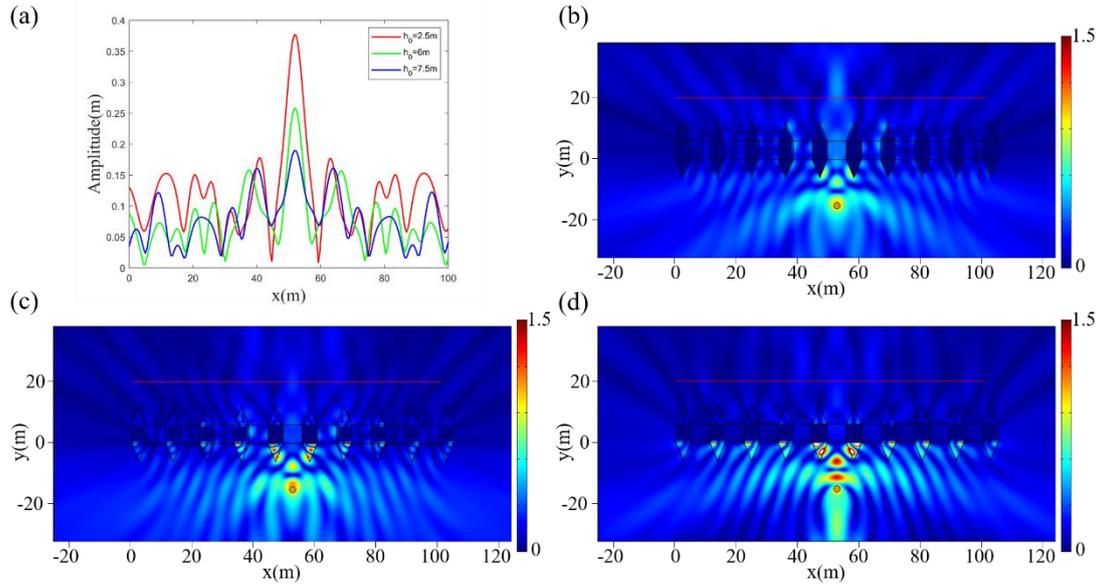

Fig. 5. Wave amplitude with a point source at (52m, -15 m). (a) The amplitude at y = 20 m for different water level $h_0$ = 2.5, 6, and 7.5 meters with the incident wavelength of 9m. (b), (c) and (d) are the field patterns for Case 1, 2, and 3, respectively.

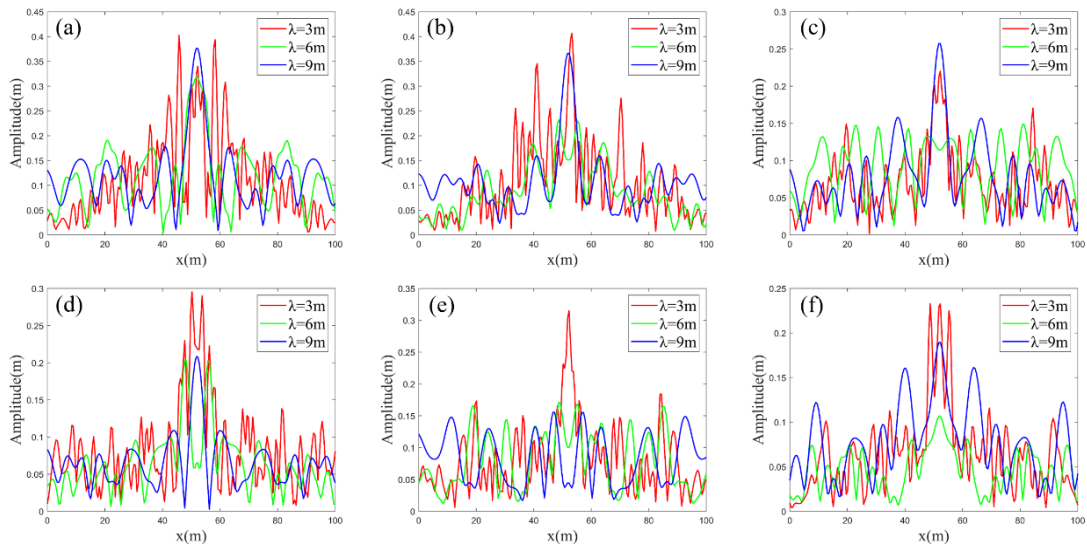

Fig. 6. The amplitude at y = 20 m for the incident wavelength $\lambda$ = 3, 6, and 9 meters at different water level $h_0$: (a) 2.5, (b) 5.5, (c) 6, (d) 6.5, (e) 7 and (f) 7.5 meters, respectively.

In conclusion, we study how the series of boat-shape pillars can reduce the water wave transmission to avoid suffering from flood from a simple metagrating model in optics. The triangular ends of the boat shaped pillars can focus waves and dissipate them effectively. In addition for further higher water lever, the bridge part between pillars severs as a zero index material, which makes the waves totally reflected from transmission. The optics theory can greatly benefit the explanation of some interesting water wave phenomena. In future, it would also be good to see demonstrations about such ancient wisdom, even with nonlinear dispersion [10] and bottom flow considered [14].

**Acknowledgments**

H. Y. C. conceived the idea and would like to thank the inspired discussions from Yixin Chen and Prof. Guirong Xie. This work is financially supported by the National Natural Science Foundation of China (Grants No. 11874311).